\documentclass[12pt]{article}
\usepackage{a4wide, amsmath, latexsym, epsfig,
  psfrag,graphicx, rotating, fancyhdr, tabularx, afterpage,booktabs}
\usepackage{subfigure}
\usepackage{relsize}
\begin{document} 
\begin{center} 
{\large \bf Study of predominant hadronic modes of  $\tau$-lepton using a Monte Carlo generator TAUOLA. 
\\ } 
\vspace{5mm} O.~Shekhovtsova 

\vspace{5mm} 
{Institute of Nuclear Physics PAN ul. Radzikowskiego 152 31-342 Krakow, Poland \\ 
           Kharkov Institute of Physics and Technology   61108, Akademicheskaya,1, Kharkov, Ukraine \\ } 
\end{center} 
\vspace{5mm} 

\begin{abstract}
TAUOLA is a Monte Carlo generator dedicated to generating tau-lepton decays and it 
is used
in the analysis of experimental data both at B-factories
and LHC.  TAUOLA is a long-term project that started in the 90's and has
been
under development up to now. 
In this note we discuss the status of the predominant hadronic tau-lepton decays into two ($Br \simeq 25.52\%$) and three pions ($Br \simeq 18.67\%$). 
\end{abstract}

\section{Introduction}
The precise experimental data for tau lepton decays collected at B-factories (both BELLE~\cite{Fujikawa:2008ma} and BABAR~\cite{Nugent:2013ij}) provide an opportunity to measure the Standard Model (SM) parameters, such as the strong coupling constant, the quark-mixing matrix, the strange quark mass etc, and for searching new physics beyond SM. The leptonic decay modes of the tau lepton allow to test the universality of the lepton couplings to the gauge bosons. Its hadronic decays (the tau lepton, due to its high mass, is the only one that can decay into hadrons) give information about the hadronization mechanism and resonance dynamics in the energy region where the methods of the perturbative QCD cannot be applied. Also hadronic flavour-violating and CP violating decays of tau lepton allow to search for new physics scenarios.

Hadronic tau lepton decays are also a tool in high-energy physics. In the last two years tau physics has received great attention after the discovery of the Higgs boson by the CMS and ATLAS  Collaborations \cite{LHC1,LHC2}. While this particle resembles in most features the SM Higgs boson, the data, showing an enhancement of the two-photon event rates, as well as a reduced rate into tau and W pairs, could indicate extended symmetries.  Another yet unsolved question is related to the origin and precise nature of  violation of symmetry under charge and parity transformations, the so called CP violation problem.

Since the 90's  TAUOLA is the main Monte Carlo (MC) generator to simulate tau-lepton decays \cite{Jadach:1993hs}. It has been used by the collaborations ALEPH \cite{Buskulic:1995ty}, CLEO  \cite{Asner:1999kj}, at both B-factories (BABAR \cite{Nugent:2013ij} and BELLE \cite{Fujikawa:2008ma}) as well at LHC \cite{LHC1,LHC2} experiments~\footnote{At LHC, at the moment, tau decays are only used for identification and are not used for studying their dynamics. However, the dynamics of tau decays are important for both modeling the decays (and therefore the reconstruction and identification) and for measuring the decay polarization.} for tau decay data analysis. The library TAUOLA can be easily attached to any MC describing the production mechanism like KORALB, KORALZ and KKMC \cite{koral1,koral2,koral3}. The code provides the full topology of the final particles including their spin.
 
In view of the forthcoming BELLE-II project \cite{Abe:2010gxa} it is important to revise the TAUOLA context in detail and we start our discussion by reviewing the models describing the tau-lepton decays into two ($Br \simeq 25.52\%$) and three pions ($Br \simeq 18.67\%$). 

\section{Two-pion and three-pion hadronic currents}

For $\tau$ decay channels with two mesons, $h_1(p_1)$ and $h_2(p_2)$, the hadronic current reads
\begin{equation}
J^\mu  = N \bigl[ (p_1 - p_2)^\mu F^{V}(s) + (p_1 + p_2)^\mu F^{S}(s) \bigr],
\end{equation}
where $s = (p_1 +p_2)^2$, $F^{V}(s)$ and $F^{S}(s)$ are the vector and scalar form factors, respectively.

In the general case, a hadronic current of a two-meson tau-lepton decay mode depends on both vector ($F^{V}$) and scalar ($F^{S}$) form factors.
However, in the isospin  symmetry limit, $m_{\pi^-} = m_{\pi^0}$, the scalar form factor vanishes for the two-pion decay mode and the current is described by the vector form factor only. The normalization coefficient ($N$) depends on the mode and equals 1 for the two-pion mode.

Currently MC TAUOLA includes four parametrizations for the vector form factor of two pions $F^{V}_{\pi}(s)$: 
\begin{itemize} 
\item Kuhn-Santamaria (KS) parametrization~\cite{Kuhn:1990ad}:
\begin{eqnarray*}
F^{V}_{\pi}(s) &=& \frac{1}{1+\beta+\gamma}(BW_\rho(s) +\beta BW_{\rho'}(s)+\gamma BW_{\rho''}(s)) \, \\ BW(s) &=& \frac{M^2}{M^2-s-i\sqrt{s}\Gamma_{\pi\pi}(s)} \, ,
\end{eqnarray*} 
where $M$ is a resonance mass and $\Gamma_{\pi\pi}(s)$ is the resonance energy-dependent width  that takes into account two-pion loops;

\item Gounaris-Sakurai (GS) parametrization used by BELLE~\cite{Fujikawa:2008ma}, ALEPH and CLEO collaborations: 
\begin{eqnarray*}
F^{V}_{\pi}(s) &=& \frac{1}{1+\beta+\gamma}(BW_\rho^{GS}(s) +\beta BW_{\rho'}^{GS}(s)+\gamma BW_{\rho''}^{GS}(s)) \, , \\ BW^{GS}(s) &=& \frac{M^2+d M\Gamma_{\pi\pi}(s)}{M^2-s+f(s)-i\sqrt{s}\Gamma_{\pi\pi}(s)} \, , 
\end{eqnarray*} 
where $f(s)$ includes the real part of the two-pion loop function;

\item parametrization based on the Resonance Chiral Lagrangian (RChL)~\cite{SanzCillero:2002bs}:
\begin{equation}
F^{V}_{\pi}(s) = \frac{1+\displaystyle\mathlarger{\sum}\limits_{i = \rho, \rho', \rho''}\frac{F_{V_i}G_{V_i}}{F^2}\frac{s}{M^2_{i}-s}}
                     {1+\left(1+\displaystyle\mathlarger{\sum}\limits_{i = \rho, \rho',\rho''} \frac{ 2G^2_{V_i}}{F^2}\frac{s}{M^2_{i}-s}\right)\displaystyle\frac{2s}{F^2}\left[B_{22}^{\pi}(s)+\frac{1}{2}B_{22}^{K}(s)\right]} \, , \nonumber
\end{equation}
where $B_{22}$ is the two-meson loop function~\footnote{Comparing the imaginary part of the loop function $B_{22}^{(\pi)}$, Eq.(A.3) in Ref.~\cite{SanzCillero:2002bs}, and Eq. (13) in~\cite{Fujikawa:2008ma} one gets $\sqrt{s}\Gamma_{\pi\pi}(s)= s\sqrt{M_V}\Gamma_V \displaystyle\frac{{\rm Im} B_{22}^\pi(s)}{{\rm Im} B_{22}^\pi(M_V)}$ for $s>(m_{\pi^-}+m_{\pi^0})^2$, where $M_i$ and $\Gamma_i$ are the resonance mass and width, respectively.}. For the physical meaning of the model parameters $F_{V_i}$ and $G_{V_i}$ see~\cite{Ecker:1988te,SanzCillero:2002bs};

\item combined parametrization (combRChL) that applies dispersion approximation at low energy and modified RChL result at high energy~\cite{Dumm:2013zh}:
\begin{eqnarray*}
s < s_0 \,:  \; \; \; \; F^{V}_{\pi}(s) &=& \exp\left[\alpha_1 s + \frac{\alpha_2}{2}s^2 + \frac{s^3}{\pi}\int\limits_{4m_\pi^2}^{\infty} ds'\frac{\delta_1^1(s')}{(s')^3(s'-s-i\epsilon)} \right] \, ,   \\
s > s_0 \, : \; \; \; \; F^{V}_{\pi}(s) &=& \frac{M_\rho^2 +(\beta + \gamma)s}{M_\rho^2 -s+\displaystyle\frac{2s}{F^2}M_\rho^2\left[B_{22}^{\pi}(s)+\frac{1}{2}B_{22}^{K}(s)\right]}
   \\
&-&\frac{\beta s}{M^2_{\rho'}-s + \displaystyle\frac{192\pi s\Gamma_{\rho'}}{M_{\rho'}\sigma_\pi^3}B_{22}^\pi(s)} -
\frac{\gamma s}{M^2_{\rho''}-s + \displaystyle\frac{192\pi s\Gamma_{\rho''}}{M_{\rho''}\sigma_\pi^3}B_{22}^\pi(s)}  ,
\end{eqnarray*} 
where $s_0$ is the high energy limit of the dispersion representation applicability. It is supposed to satisfy $1.0$GeV$^2 < s_0 < 1.5$GeV$^2$~\cite{Dumm:2013zh} and we leave it to be fitted. 
\end{itemize}  

In all the above parametrizations, except for the RChL one, the pion form factor is given by interfering amplitudes from the known isovector meson resonances $\rho(770)$, $\rho'(1450)$ and $\rho''(1700)$ with relative strengths $1$, $\beta$ and $\gamma$.  Although one could expect from the quark model that $\beta$ and $\gamma$ are real, we allow the parameters to be complex (following the BELLE, CLEO and ALEPH analysis)  and their phases are left free in the fits. 
In the case of the RChL parametrization we restrict ourselves to  the $\rho(700)$ and $\rho'(1450)$ contributions, with the relative $\rho'$ strength  (that is a combination of the model parameters $F_{V_i}$, $G_{V_i}$ and $F$) being a real parameter. 
 
For the energy-dependent width  of $\rho(770)$-meson two-pion and two-kaon loop contributions are included for both  RChL and combRChl parametrizations through the loop functions $B_{22}^{\pi}$ and $B_{22}^{K}$, whereas in the case of KS and GS the $\rho$ width is approximated only by the two-pion loops through $\Gamma_{\pi\pi}$. 
The $\rho'(1450)$ and $\rho''(1700)$ widths include only two-pion loops for all parametrizations except for the RChL one. In the case of the RChL one both two-pion and two-kaon loops are included.

Results of the fit to the BELLE data \cite{Fujikawa:2008ma} are presented in Figs.~1 and 2. The best fit is with the GS pion form factor ($\chi^2 = 95.65 $) and the worst one is with the RChL one ($\chi^2 = 156.93 $), which is not able to reproduce the high energy tail. As it was mentioned above, the two main differences of the RChL parametrization compared to the others are 1) the $\rho''$-meson absence, 2) a real value of the $\rho'(1450)$-meson strength. To check the influence of the $\rho''(1700)$ on the RChL result, the $\rho''(1700)$ resonance has been included in the same way as it was done for $\rho'(1450)$; however, this inclusion has not improved the result (see, Fig. 2, the right panel).

In an effective field theory, like RChL, complex values come only from loops.
Therefore, we conclude that missing loop contributions could be responsible for the disagreement and that the complex value of the $\beta$ and $\gamma$ parameters 
might mimic  missing multiparticle loop contributions. The same conclusion was reached in Ref.~\cite{SanzCillero:2002bs} where it was stressed that the two-pseudoscalar loops cannot incorporate all the inelasticity needed
to describe the data and other multiparticle intermediate states can play a role.   
We are going to check this point by adding first a four-pion loop contribution
 to the $\rho'$-resonance propagator.
\begin{figure}[h]
\includegraphics[scale = .4]{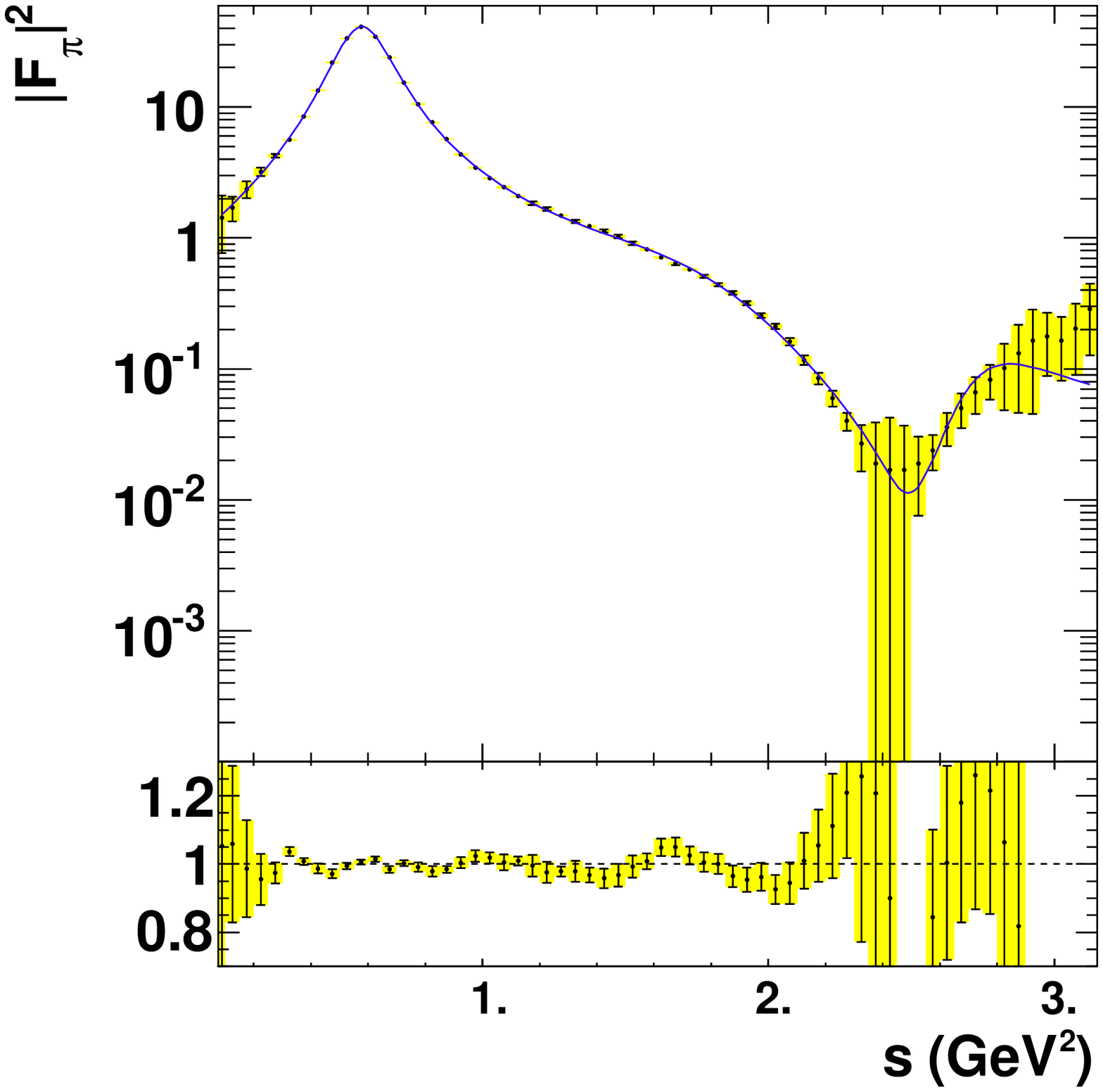}
\includegraphics[scale = .4]{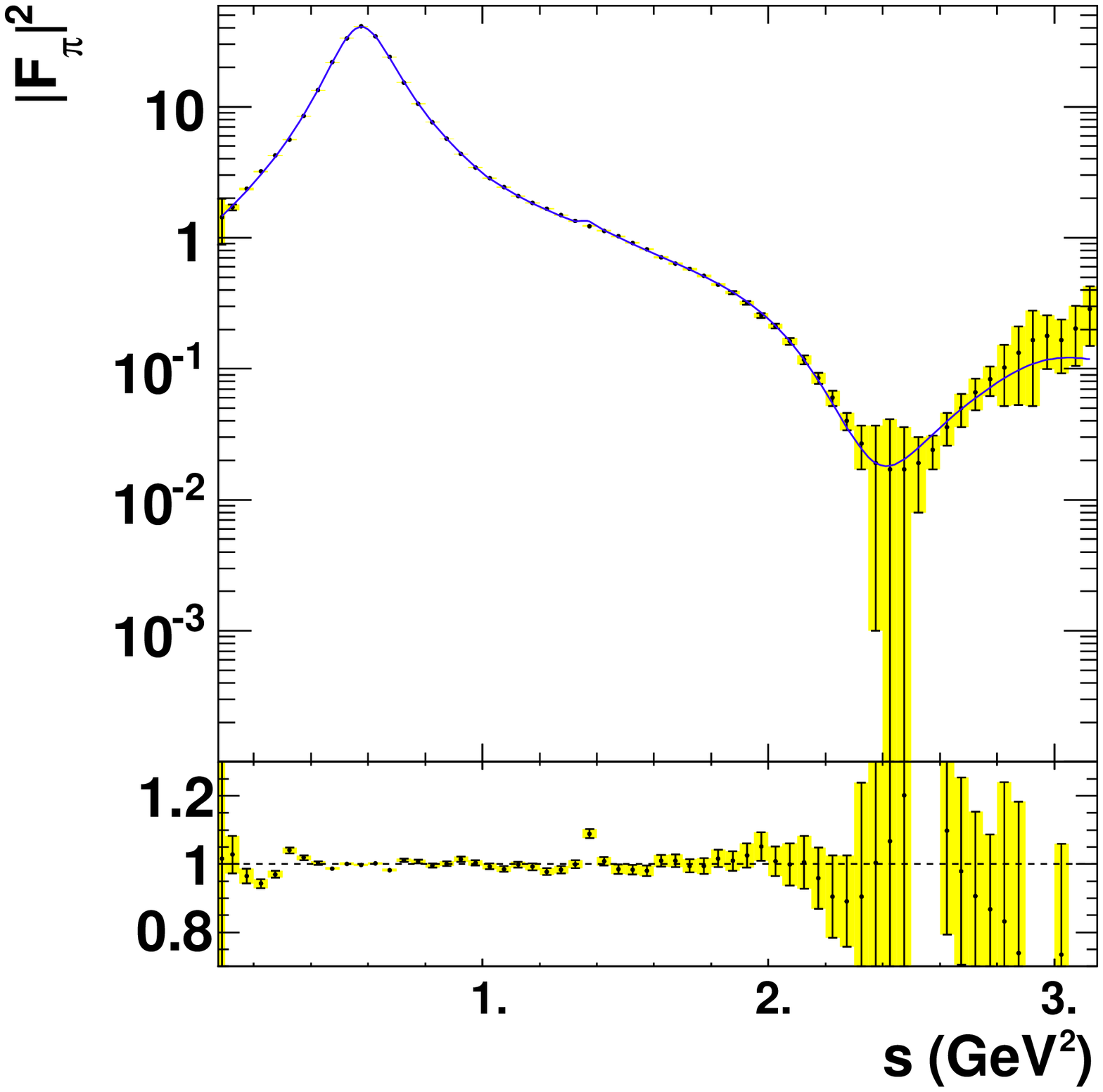}
\caption{The pion form factor fit to Belle data \cite{Fujikawa:2008ma}: the GS parametrization (left panel), the RChL parametrization (right panel). At the bottom of the figure, the ratio of a theoretical prediction
to the data is given.}
\end{figure}

\begin{figure}[h]
\includegraphics[scale = .4]{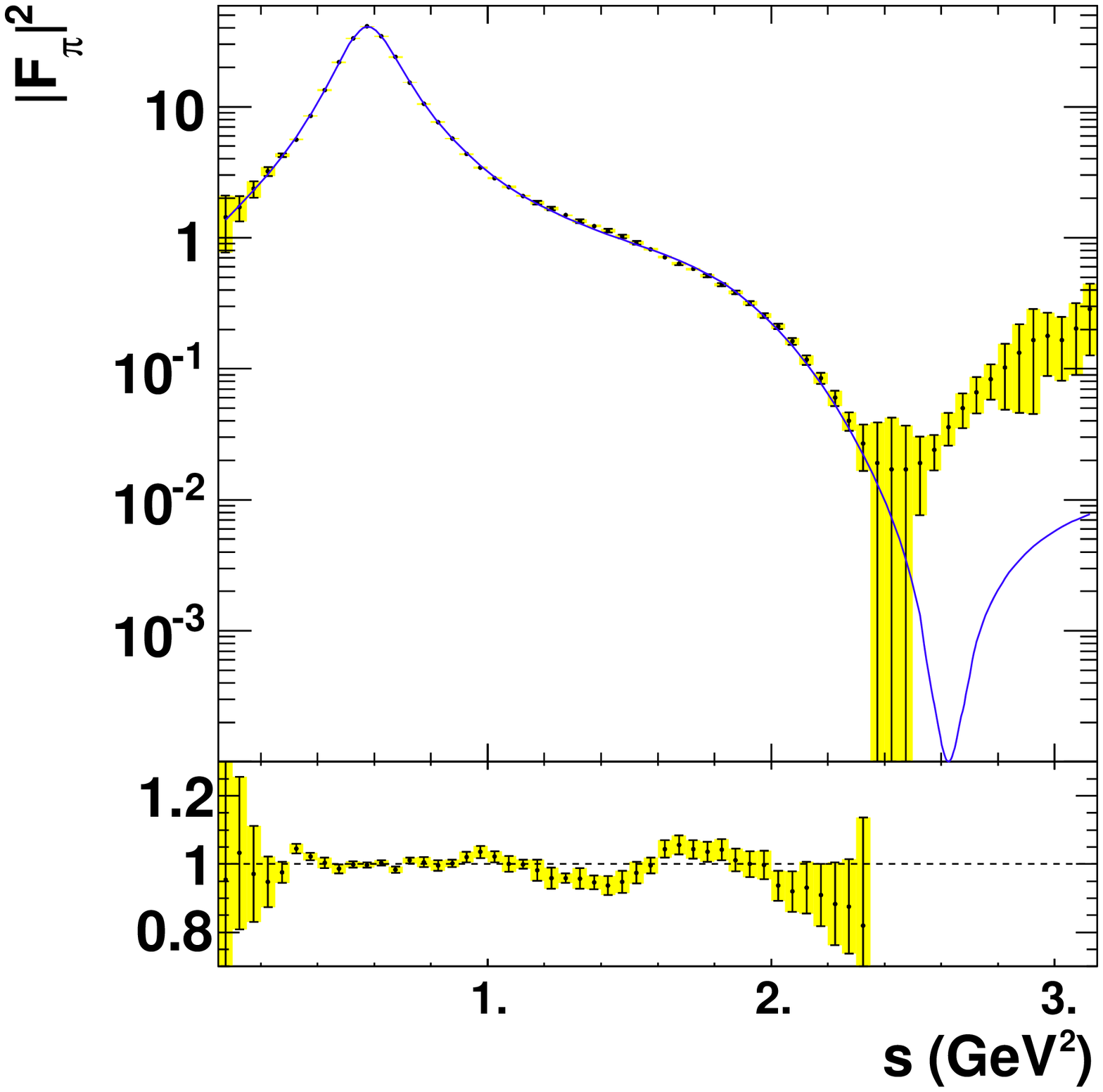}
\includegraphics[scale = .4]{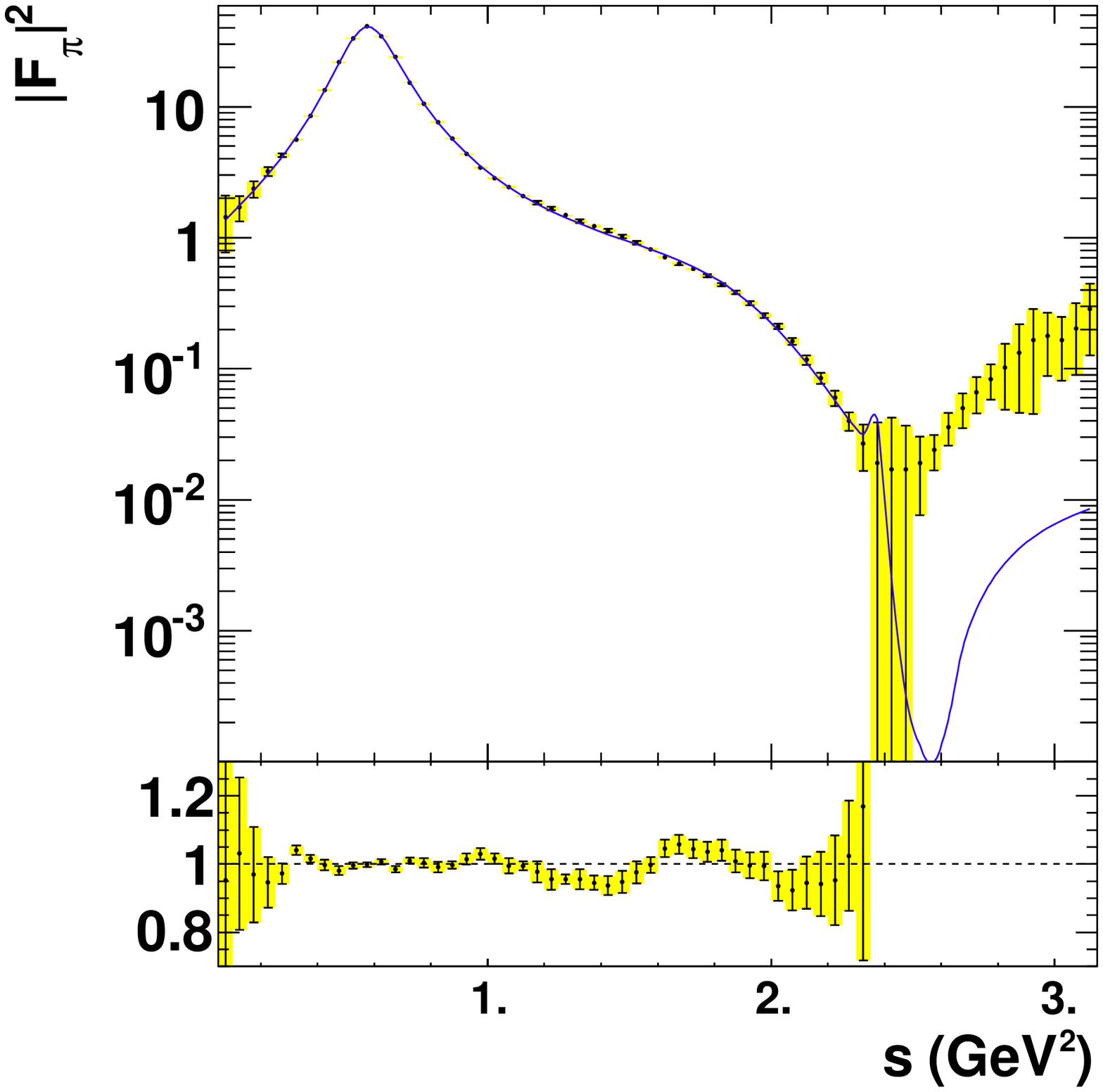}
\caption{The pion form factor within the comRChL parametrization is fitted to BELLE data \cite{Fujikawa:2008ma}: the fit without $\rho''(1700)$ (left panel) and with it (right panel). At the bottom of the figure, the ratio of a theoretical prediction
to the data is given.}
\end{figure}

In spite of the fact that the GS parametrization is able to reproduce the BELLE data, it has at least two misleading points. One of them is the complex strengths mentioned above, the other is the neglect of the two-kaon loops even for the $\rho$-resonance propagator. The last point will be the object of the further investigation.

In the case of the combRChL parametrization the fitting curvature demonstrates not a smooth behaviour near $s = s_0$. Therefore  more sophisticated fitting techniques will have to be implemented.

When the $\tau$ lepton decays into three hadrons and a neutrino, the predominant decay mode involves three pions. In the general case Lorentz invariance determines the decomposition of the hadronic current for a three-hadron final state in  terms of  five Lorentz invariant structures \cite{Shekhovtsova:2012ra} multiplied by hadronic form factors ($F_i$)
\begin{eqnarray}
J^\mu &=N &\bigl\{T^\mu_\nu \bigl[ c_1 (p_2-p_3)^\nu F_1(q^2,s_1,s_2)  + c_2 (p_3-p_1)^\nu
 F_2(q^2,s_1,s_2)  + c_3  (p_1-p_2)^\nu F_3(q^2,s_1,s_2) \bigr]\nonumber\\
& & + c_4  q^\mu F_4(q^2,s_1,s_2)  -{ i \over 4 \pi^2 F^2}      c_5
\epsilon^\mu_{.\ \nu\rho\sigma} p_1^\nu p_2^\rho p_3^\sigma F_5(q^2,s_1,s_2)      \bigr\},
\label{fiveF}
\end{eqnarray}
where as usual  $T_{\mu\nu} = g_{\mu\nu} - q_\mu q_\nu/q^2$ denotes the transverse
projector, and $q^\mu=(p_1+p_2+p_3)^\mu$ is the total momentum of the hadronic system with four-momenta $p_1$, $p_2$ and $p_3$ and the two-pion invariant mass squared is $s_i = (p_j + p_k)^2$.

The scalar functions $F_i(q^2,s_1,s_2)$ are the three-pion form factors. The vector form factor vanishes for the three-pion modes due to the
G-parity conservation $F_5 = 0$. Among the three hadronic form factors $F_i$, $i = 1, 2, 3$  which correspond 
to the axial-vector part of the hadronic tensor, only two are independent. The pseudoscalar form factor $F_4$ is proportional to $m^2_\pi/q^2$~\cite{Shekhovtsova:2012ra}, thus it is suppressed with
respect to $F_i$, $i = 1, 2, 3$.

In TAUOLA the following three-pion form factors are available 
\begin{itemize} 
\item CPC version \cite{Jadach:1993hs}, that includes only the dominant $a_1 \to \rho\pi$ mechanism production. The form factor is a product of the Breit-Wigner for the $a_1$ and $\rho$ meson;
  
\item CLEO parametrization;
It is based on the Dalitz plot analysis carried out by the CLEO collaboration and includes the following intermediate states: $a_1 \to (\rho;\rho')\pi$ (where $(\rho;\rho')$ can be either in S- or in D-wave), $a_1 \to \sigma\pi$, $a_1 \to f_2(1270)\pi$, $a_1 \to f_0(1370)\pi$.   
In fact, there are two variants of this parametrization. The former is based on the CLEO  $\pi^0\pi^0\pi^-$  analysis  \cite{Asner:1999kj} and  applies the same current for the $\pi^-\pi^-\pi^+$ mode.
 The later uses the $\pi^0\pi^0\pi^-$ current from \cite{Asner:1999kj} and the $\pi^-\pi^-\pi^+$ current from  the unpublished CLEO analysis~\cite{Shibata:2002uv}. It is interesting to point out that a difference between these variants of the CLEO parametrization is related with the scalar and tensor resonance contributions. All resonances are modelled by Breit-Wigner functions and the hadronic current is a weighted sum of their product~\footnote{This approach was contested in Ref.~\cite{GomezDumm:2003ku} where it was demonstrated that the corresponding hadronic form factors reproduced the leading-order chiral result and failed to reproduce
the next-to-leading-order one.}. The model parameters are resonance masses and widths as well as the weights;

\item modified RChL parametrization \cite{Nugent:2013hxa}. 
It was based on the RChl results for the three-pion currents~\cite{Dumm:2009va} and an additional scalar resonance contribution. The RChL current is a sum of the chiral contribution corresponding to the direct vertex $W^- \to \pi\pi\pi$, single-resonance contributions, e.g. $W^- \to \rho \pi$, double-resonance contributions, as $W^- \to a_1^- \to \rho \pi$. Only vector and axial-vector are included in the RChL hadronic currents. More details about the parametrization can be found in~\cite{Shekhovtsova:2012ra}.
The scalar resonance contribution was included phenomenologically by requiring the RChL structure for the currents and modelling the $\sigma$-resonance by a  Breit-Wigner function.
\end{itemize} 

The CLEO parametrization for the $\pi^-\pi^-\pi^+$ mode has not yet been fitted to the BABAR preliminary data \cite{Nugent:2013ij}, so we used the old numerical values of the parameters fitted to  the CLEO data~\cite{Asner:1999kj}. The fit to the BABAR data will be a task for future work.
\begin{figure}[h]
\includegraphics[scale = .27]{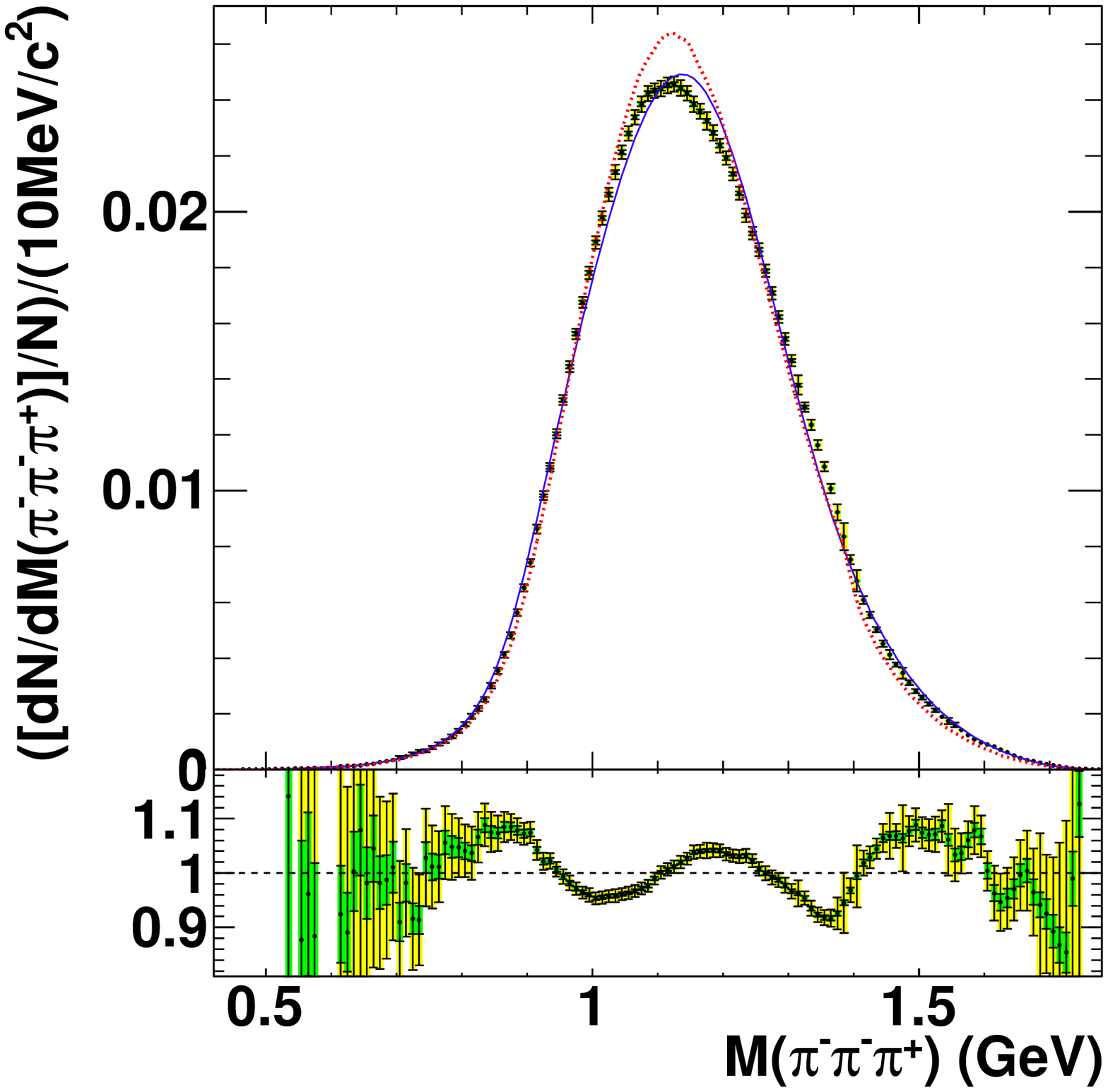}
\includegraphics[scale = .27]{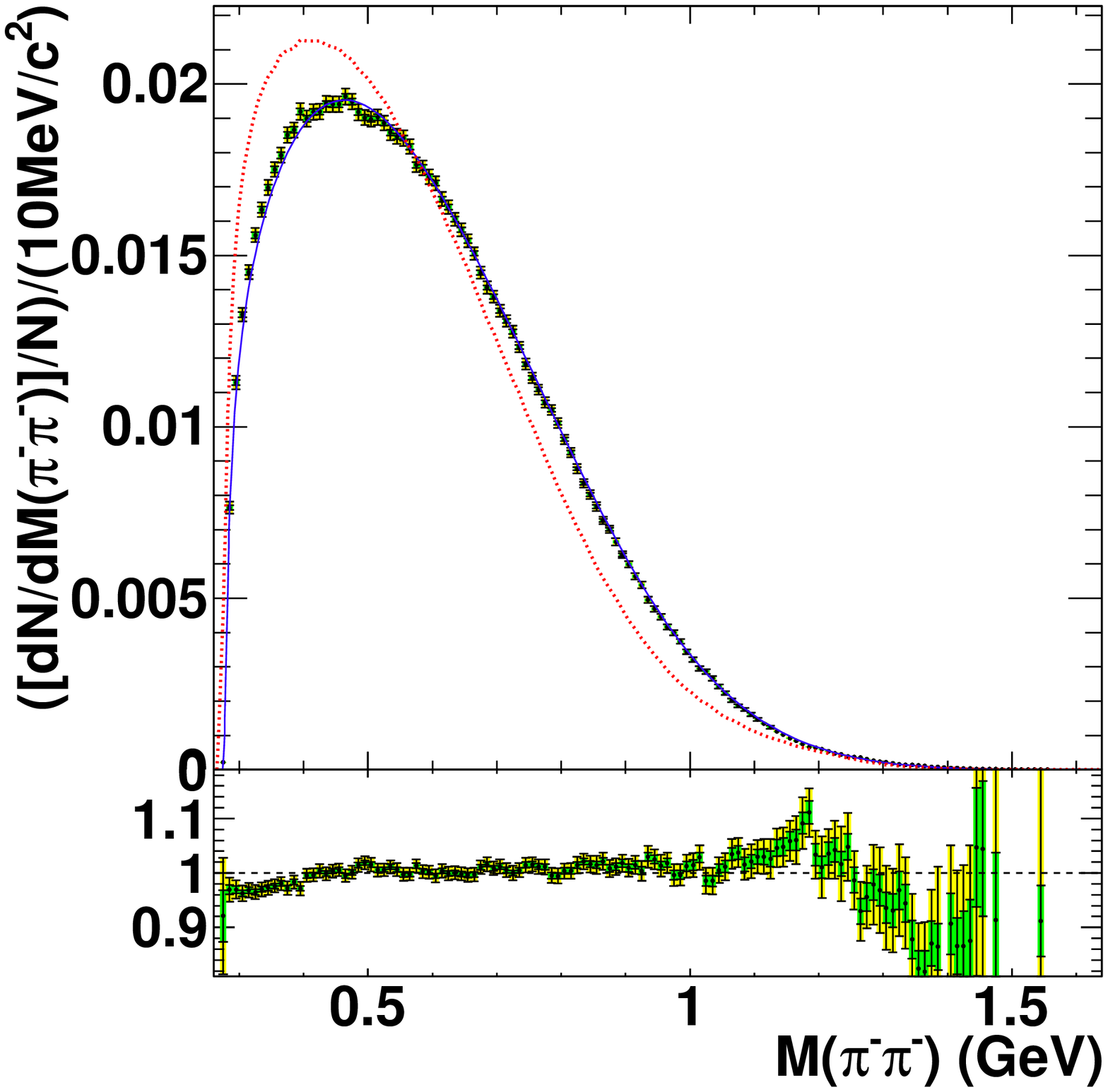}
\includegraphics[scale = .27]{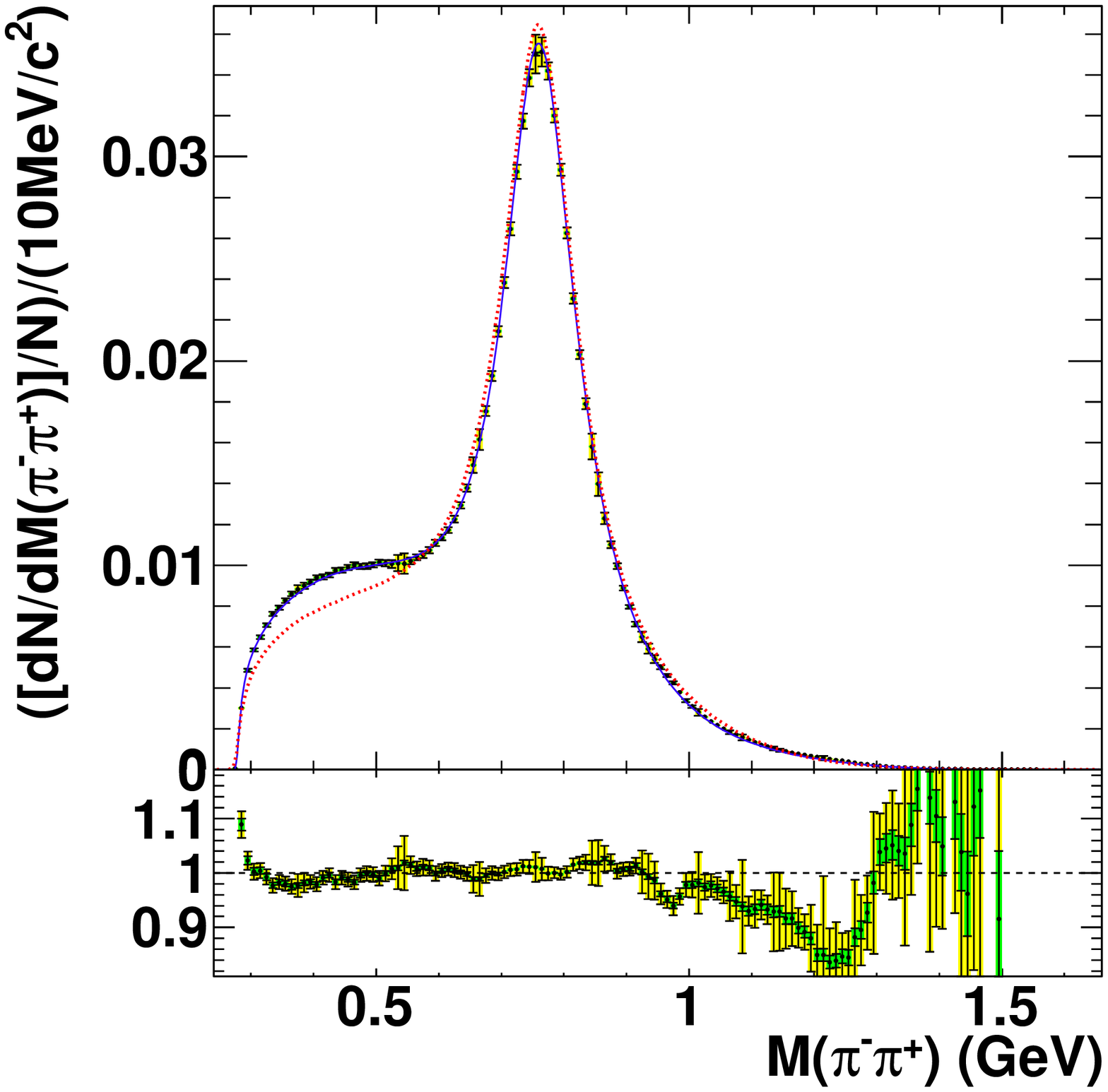}
\caption{The $\tau^-\to \pi^-\pi^-\pi^+\nu_\tau$ decay invariant mass distribution of the three-pion system (left panel) and the two-pion system (central and right panels).
The BABAR data~\cite{Nugent:2013ij} are represented by the data points, with the results from the modified RChL
current as described in the text (blue line) and the old tune from CLEO~\cite{Asner:1999kj} 
(red-dashed line) overlaid.
}
\end{figure}

The one-dimensional distributions of the two- and three-pion invariant mass spectra calculated on the base of the modified RChL parametrization have been fitted to the BABAR preliminary data~\cite{Nugent:2013ij}. It must be pointed out that without the scalar resonance contribution the RChL parametrization provides a slightly better result than the CLEO parametrization whereas the scalar resonance inclusion strongly improves the low two-pion mass invariant spectrum.
Discrepancy between  theoretical spectra and experimental data can be explained by missing resonances in the model, such as the axial-vector resonance $a_1'(1600)$, the scalar resonance $f_0(980)$ and the tensor resonance $f_2(1270)$. Inclusion of these resonances in the RChL framework will be a future task. 

Comparison of the $\pi^-\pi^-\pi^+$ current in the framework of the modified RChL with the ChPT result has demonstrated that the scalar resonance contribution has to be corrected to reproduce the low energy ChPT limit. The corresponding calculation is in progress.

\section{Conclusion}

In this note we have reviewed the physical frame of the Monte Carlo generator TAUOLA for the part related with two- and three-pion decay modes. We have briefly described the models used for the hadronic currents as well as the results of the preliminary fits to the experimental data. 

In the case of the two-pion modes we have made a fit to the BELLE data for the three parametrizations of the pion form factor. We have reproduced the data with the Gounaris-Sakurai pion form factor parametrizations while have failed with the RChL parametrization. We have concluded that the complex value of the resonance strength, used in the Gounaris-Sakurai parametrization, could mimic the missing multiparticle loops. This idea will be tested by adding the four-pion loops to the $\rho'$-resonance propagator that will be the object of future study.   

In the case of the three-pion mode TAUOLA has two parametrizations for the hadronic current. At present we have fitted the modified Resonance Chiral current, which includes the $\sigma$-meson contribution in addition to the Resonance Chiral Lagrangian current, to the preliminary BABAR data. We have demonstrated that the difference between the model and the data is within $5-7\%$ at the high-tail of the two- and three-pion spectra. This discrepancy is related to the neglect of the resonances, like $f_0(980)$, $a_1'(1600)$ and $f_2(1270)$, in the model. The incorporation of these resonances is in progress.   

The same type of comparison will have to be done for the CLEO $\pi^-\pi^-\pi^+$ current. 

One of the key points of the BELLE-II project mentioned in Introduction is tau physics~\cite{Abe:2010gxa,Bevan:2014iga}. The first physics run is planned in the fall of 2018.  Both allowed and forbidden  tau decay modes in  Standard Model will be measured. 
It should record a 50 times larger data sample than the BELLE experiment until 2022. 
Indeed, the two- and three-pion decay modes will be used for measuring resonance parameters on one side and on the other these processes constitute an ever present background to the hadronic decay modes, forbidden in the Standard Model, for example the lepton flavour violated modes, like $\tau \to l \pi\pi$~\cite{Celis:2013xja}.
All these considerations underline the importance of a precise theoretical description of the two- and three-pion decay modes.

\section*{Acknowledgements}

This research was supported in part from funds of Foundation of Polish Science grant POMOST/2013-7/12, that is co-financed from European Union, Regional 
Development 
Fund, and by Polish National Science 
Centre under decisions  DEC-2011/03/B/ST2/00107.

\end{document}